\documentclass[sigconf]{acmart}%
\AtBeginDocument{%
  \providecommand\BibTeX{{%
    \normalfont B\kern-0.5em{\scshape i\kern-0.25em b}\kern-0.8em\TeX}}}

\copyrightyear{2023}
\acmYear{2023}
\setcopyright{acmlicensed}\acmConference[UIST '23]{The 36th Annual ACM Symposium on User Interface Software and Technology}{October 29-November 1, 2023}{San Francisco, CA, USA}
\acmBooktitle{The 36th Annual ACM Symposium on User Interface Software and Technology (UIST '23), October 29-November 1, 2023, San Francisco, CA, USA}
\acmPrice{15.00}
\acmDOI{10.1145/3586183.3606804}
\acmISBN{979-8-4007-0132-0/23/10}

\usepackage{siunitx}

\begin{document}

\title{3D Printing Magnetophoretic Displays}

\author{Zeyu Yan}
\email{zeyuy@umd.edu}
\affiliation{%
  \institution{University of Maryland}
  \city{College Park}
  \state{Maryland}
  \country{USA}
  \postcode{20740}
}

\author{Hsuanling Lee}
\email{lee3050@purdue.edu}
\affiliation{%
  \institution{Purdue University}
  \city{West Lafayette}
  \state{Indiana}
  \country{USA}
  }

\author{Liang He}
\email{lianghe@purdue.edu}
\affiliation{%
  \institution{Purdue University}
  \city{West Lafayette}
  \state{Indiana}
  \country{USA}
  }

\author{Huaishu Peng}
\email{huaishu@umd.edu}
\affiliation{%
  \institution{University of Maryland}
  \city{College Park}
  \state{Maryland}
  \country{USA}
  \postcode{20740}
}

\renewcommand{\shortauthors}{Yan, et al.}

\begin{abstract}
We present a pipeline for printing interactive and always-on magnetophoretic displays using affordable Fused Deposition Modeling (FDM) 3D printers. Using our pipeline, an end-user can convert the surface of a 3D shape into a matrix of voxels. The generated model can be sent to an FDM 3D printer equipped with an additional syringe-based injector. During the printing process, an oil and iron powder-based liquid mixture is injected into each voxel cell, allowing the appearance of the once-printed object to be editable with external magnetic sources. To achieve this, we made modifications to the 3D printer hardware and the firmware. We also developed a 3D editor to prepare printable models. We demonstrate our pipeline with a variety of examples, including a printed Stanford bunny with customizable appearances, a small espresso mug that can be used as a post-it note surface, a board game figurine with a computationally updated display, and a collection of flexible wearable accessories with editable visuals.
\end{abstract}



\begin{teaserfigure}
  \includegraphics[width=\textwidth]{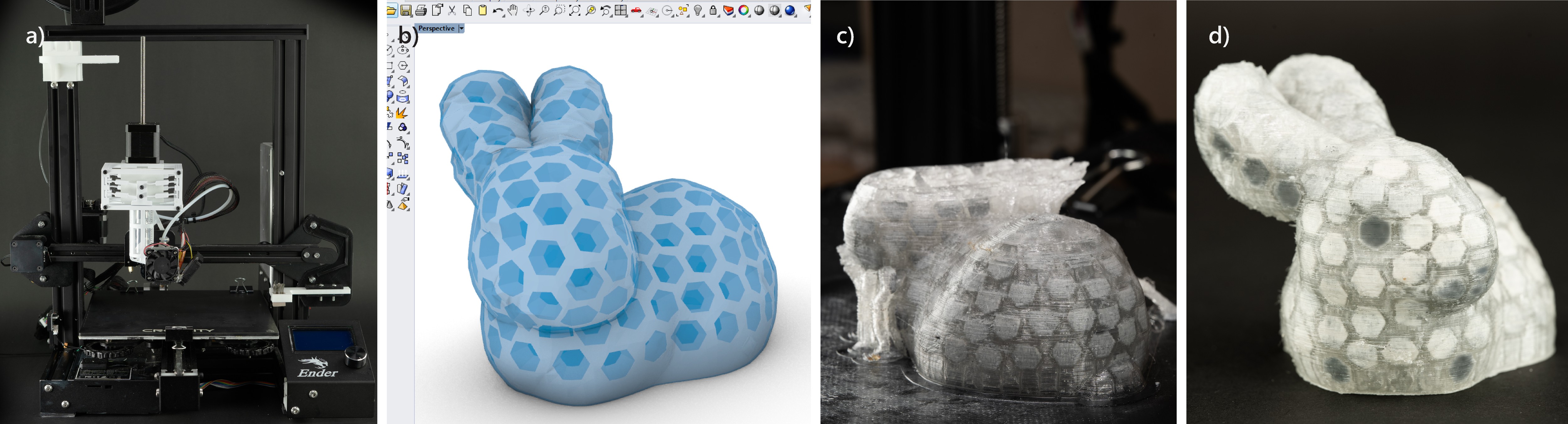}
  \caption{Printing pipeline overview. a) A modified FDM 3D printer with an additional syringe injector. b) A 3D editor that converts a model into a magnetophoretic display. c) Printing in-progress. d) The appearance of the printed model can be post edited.}
  \Description{Printing pipeline overview. a) A modified FDM 3D printer with an additional syringe injector. b) 3D editor that converts a model into a magnetophoretic display. c) Printing in-progress. d) The appearance of the printed model can be post edited.}
  \label{fig:teaser}
\end{teaserfigure}

\maketitle

\section{Introduction}
Research advancements in 3D printing have enabled end users to design and fabricate a variety of 3D dynamic artifacts. For example, research has shown that 3D printed bespoke objects can deform~\cite{he2022kinergy,he_ondule}, produce sound~\cite{ishiguro20143d}, and exhibit a range of mechanical behaviors~\cite{peng_3d_2016,lin_flexhaptics}. However, printing 3D artifacts with appearances that are non-static or interactive remains challenging.

Recent studies, such as Printed Optics~\cite{willis2012printed},  PAPILLON~\cite{brockmeyer2013papillon}, and computational light routing~\cite{pereira2014computational}, propose to alter the appearance of a 3D printed artifact by printing embedded transparent pipes that redirect images and lights from a 2D digital display. Since the optic pipes must be printed with high resolution in the direction of pipe growth, high-end inkjet or PolyJet printing methods are required. The need for additional flat displays at the base of the models also limits the type of objects that can be printed (require a large base area), and prevents them from functioning as always-on displays. 

In this paper, we present an end-to-end pipeline that converts free-form custom artifacts to be self-contained, always-on, and interactive 3D surface using a modified consumer-grade FDM 3D printer. Our technique is greatly inspired by the widely used 2D magnetophoretic displays, also known as Magna Doodles, which are commonly seen as drawing pads for children and in applications such as underwater whiteboards while scuba diving. The 2D magnetophoretic display is filled with a mixture of white liquid and dark soft-iron particles that are encapsulated within small hexagonal cells and react to external magnetic forces, causing the displayed appearance to change.
The altered appearance will remain unchanged without power, until a new external stimulus appears.
Our goal is to bring such 2D magnetophoretic displays into the third dimension using affordable FDM printers. Specifically, we aim to create 3D custom objects whose appearance can be altered and reprogrammed after the prints.  

Our pipeline consists of a modified FDM 3D printer with an additional syringe-type injector and a 3D design editor that converts a 3D model's surface into a matrix of magnetophoretic cells.
In the remainder of this paper, we will describe the modifications made to the printer and the series of experiments conducted on the printing materials and cell geometries that, all together, ensure the printing of a matrix of small cells suitable for a bi-color, always-on display.
Following the hardware modifications, we will detail the computational procedure used to convert the surface of a 3D model to the magnetophoretic cells, as well as the generation of the corresponding G-code. 
Finally, we will present a set of application scenarios that showcase our pipeline and method.

In summary, we propose a 3D printing pipeline to create free-form 3D objects with always-on and interactive appearances. Our main contributions include: 1) a low-cost FDM 3D printer hardware add-on; 2) the experiment on the materials and printing parameters; 3) an interactive 3D editor to enable the creation of objects with magnetophoretic surfaces; and 4) application scenarios that showcase the objects that can be created with our pipeline. 

\section{Related Work}
Our contribution builds upon prior personal fabrication research~\cite{mota2011rise, willis2010interactive, baudisch2017personal} that focuses on printed artifacts with custom displays, programmable magnetic fields, embedded liquid, and the computational methods that generate printable 3D meshes. 

\subsection{Fabricated Artifacts with Integrated Displays}
Much research in HCI has proposed enabling techniques to manufacture displays with multiple segments or of different shapes using electrochromic ink~\cite{jensen2019transprint}, recycled electronic ink~\cite{hanton2022fabricatink}, and electroluminescent material~\cite{klamka2017illumipaper, olberding2014printscreen}. As many of these approaches work with thin, flat substrates, these custom and flexible displays have been integrated in the user's surroundings ~\cite{olberding2014printscreen, groeger2018objectskin} as ambient displays~\cite{ishii1997tangible} or directly applied on or to the user's body ~\cite{jarusriboonchai2020always, withana2018tacttoo}, as wearables or e-tattoos. 

While the use of thin substrates enables the customization of flexible displays, the majority of them are restricted to simple 3D geometries, i.e., they cannot be easily applied to highly curved or irregular geometries. To overcome this limitation, researchers have proposed to use conductive or photochromic sprays to retrofit an existing object. 
For example, ColorMod~\cite{punpongsanon2018colormod}, Photo-Chromeleon~\cite{jin2020photo} and ChromoUpdate ~\cite{wessely2021chromoupdate} propose to spray paint multi-color photochromic inks on to the surface of 3D printed artifacts. By selectively exposing the photochromic coating to a set of UV light sources, the surface of these 3D artifacts can be programmed and updated to different colors and patterns. Similar concept can also be seen with ProtoSpray~\cite{hanton2020protospray, wessely2020sprayable}, where electroluminescent inks are used to prototype interactive displays and sensors at large scale.

Another set of research aims to incorporate display components --- light pipes~\cite{willis2012printed, brockmeyer2013papillon}, reflective diffusers~\cite{dierk2022project},  fibers~\cite{pereira2014computational,tone2020fibar}, or LEDs~\cite{torres2017illumination} --- during the process of fabrication, instead of retrofitting the object after it has been made. 
For example, Printed Optics~\cite{willis2012printed} and PAPILLON~\cite{brockmeyer2013papillon} propose to embed optical elements during 3D printing to make artifacts, such as toy figures, with custom display surfaces.  
Computational light routing~\cite{pereira2014computational} and FibAR~\cite{tone2020fibar} propose fiber design algorithm which automatically routes optical channels between two surfaces to create displays of custom shapes. 
Note that although the display side of these artifacts can have a complex geomety, the other side must be flat, as a traditional flat screen needs to be instrumented as the light source to make the printed display work. 
Recently, Zeng \textit{et.al.} ~\cite{zeng2021lenticular} proposed printing voxelized lenticular lenses across the curved surface of objects to make artifacts with different appearances when viewed from different viewpoints. 
These printed displays do not require powered light sources, but the display patterns must to be pre-programmed prior to fabrication and cannot be edited after the print has been made. 

In our paper, we also ``voxelize'' the surface of an object to make it a display. 
Unlike previous research, our display is made by injecting magnetic powder-based liquid into the object's surface. 
The printing process doesn't require expensive inkjet or PolyJet printing method, and can be used without power source.

\subsection{Fabricated Artifacts with Embedded Magnets}

In recent years, HCI researchers have attempted to include additional functionality in 3D printed objects with embedded electromagnets. 
For example, MagTics~\cite{pece_magtics_2017} and Programmable Polarities~\cite{nisser_programmable_2021} embed electromagnets into the printed objects to render haptics and create mechanical motions. 
Peng et al.'s magnetic 3D printer ~\cite{peng_3d_2016} further showcases the possibility of printing simple tangible displays combining 3D printed electromagnets and ferrofluids. 
A body of work in recent HCI reserach \cite{Magnetact, MechCircuit, Mixels} demonstrate that incorporating programmable magnetic behaviors can create unique opportunities in interactive experiences.

Permanent magnets have also been used to make 3D printed objects interactive. For example, several studies~\cite{zheng_mechamagnets_2019, ogata_magneto-haptics_2018, yasu2022magneshape, ogata_computational_2021, tasnim2020mechanobeat} show that permanent magnets can be combined with various 3D printed constraints to make custom knobs and sliders with rich haptic feedback, or create harmonic oscillators that can be used as mechanical tags for tracking daily interactions. 
Permanent magnets have also been widely used as the key connection components between 3D printed objects for instant 3D modeling~\cite{suzuki_dynablock_2018}, or rapid electronics prototyping~\cite{schmitz_oh_2021, bdeir_electronics_2009}.

Our work is inspired by previous work in that we also revolve around embedding magnetizable material inside 3D printed objects, but with the focus on changing their appearances rather than mechanical behaviors.

\subsection{Fabricated Artifacts with Integrated Fluids}

Another branch of work that we draw inspiration from are the technologies developed to integrate liquids into fabricated objects. 
In microfluidics, for example, material science and chemical engineering researchers have investigated several approaches~\cite{kitson_combining_2013, owens_high-precision_2018, duffy_rapid_1998, tothill2017fabrication} for fabricating chambers and tunnels with dimensions in the micrometer range or smaller. 
HCI research has also demonstrated different ways to embed cell-based structures can potentially be repurposed for fluid embedding \cite{InfraredTags, AirCode}.
Recent work, like Venous Materials~\cite{mor_venous_2020} and OmniFiber~\cite{kilic_afsar_omnifiber_2021}, has also explored the use of microfluidics as novel displays and actuators.

Liquid or slurry has also been explored as a new 3D printing material. For example, Off-Line Sensing~\cite{schmitz2018off} utilizes manually embedded liquids in 3D printed objects as on-demand sensors to passively detect when the objects are tilted. FabHydro~\cite{yan2021fabhydro} embeds liquid into 3D printed bellows structures to create hydraulic-driven mechanical devices.  ExpandFab~\cite{kaimoto_expandfab_2020} prints objects with foam slurry which can change its shape and volume when exposed to heat. In Unmaking~\cite{song_unmaking_2021}, researchers explore how 3D printed objects can fail and be destroyed in a creative manner using embedded microsphere slurry. 

Our work takes inspiration from the above and also embeds viscous liquid into the printed object. Because the liquid will be deposited into a matrix of small voxels, we have engineered a dual-nozzle system and customized the corresponding G-code to automate the liquid injection process.

\subsection{Computational Design of Appearance Changeable 3D Models}
To create 3D printable models that can alter the appearance, researchers have explored novel computational approaches in the 3D modeling process ~\cite{tone2020fibar, ma2017pneumatic, savage2014series}. 
One approach is to directly change the underlying structures of the 3D object's surface ~\cite{ion2018metatextures, rouiller20133d, zeng2021lenticular}. 
For example, Lenticular Objects ~\cite{zeng2021lenticular} distributes 3D printable multi-color lenses evenly on the surface of the input model to achieve the desired visual appearance at various view angles. The lens placements are calculated by remeshing the model's surface into an isotropic triangular mesh and allocating the triangle vertices as the lens's center positions. 

In addition to adding appearance-altering mechanisms to the 3D model surfaces, the model body can also be configured to change the physical appearance through computationally generated mechanisms ~\cite{ma2017pneumatic, tone2020fibar, punpongsanon2018colormod}. 
 For example, using generative Fibonacci lattices, PAPILLON~\cite{brockmeyer2013papillon} computes the pixel distributions on both the planar image source and the spherical surface of a model. The corresponding pixels of the two surfaces are connected through a 3D-printed pipe filled with clear optical materials for image transmission. Savage et al. ~\cite{savage2014series} develop an A*-based 3D path routing algorithm that optimizes the 3D printed pipes inside a model. These pipes contain electroluminescent wires that can produce light animations.

Similar to~\cite{zeng2021lenticular}, our pipeline also use Instant Meshes library~\cite{jakob2015instantmeshes} to generate 3D cells across the surface of the model. Unlike previous works, our computational procedure is optimized for generating small watertight cells that are printable with low-cost desktop FDM 3D printers.

\section{Magnetophoretic Display Principle} \label{principle}

\begin{figure}[h]
  \includegraphics[width=\columnwidth]{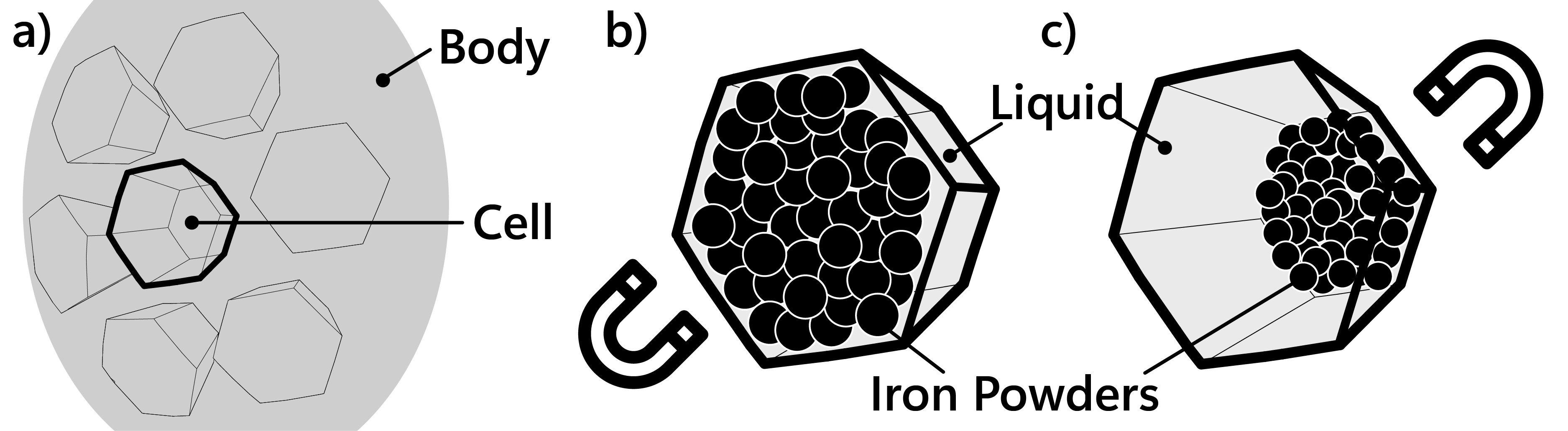}
  \caption{Magnetophoretic display principle: a) illustrations of the side view of one single cell with magnetic powder on the front side (left) and the (back side) right. b) and c) the visual effect of a magnetophoretic display cell being triggered.}
  \Description{Magnetophoretic display principle: a) illustrations of the side view of one single cell with magnetic powder on the front side (left) and the (back side) right. b) and c) the visual effect of a magnetophoretic display cell being triggered.}
  \label{fig:magDisplay}
\end{figure}

\begin{figure}[h]
  \includegraphics[width=\columnwidth]{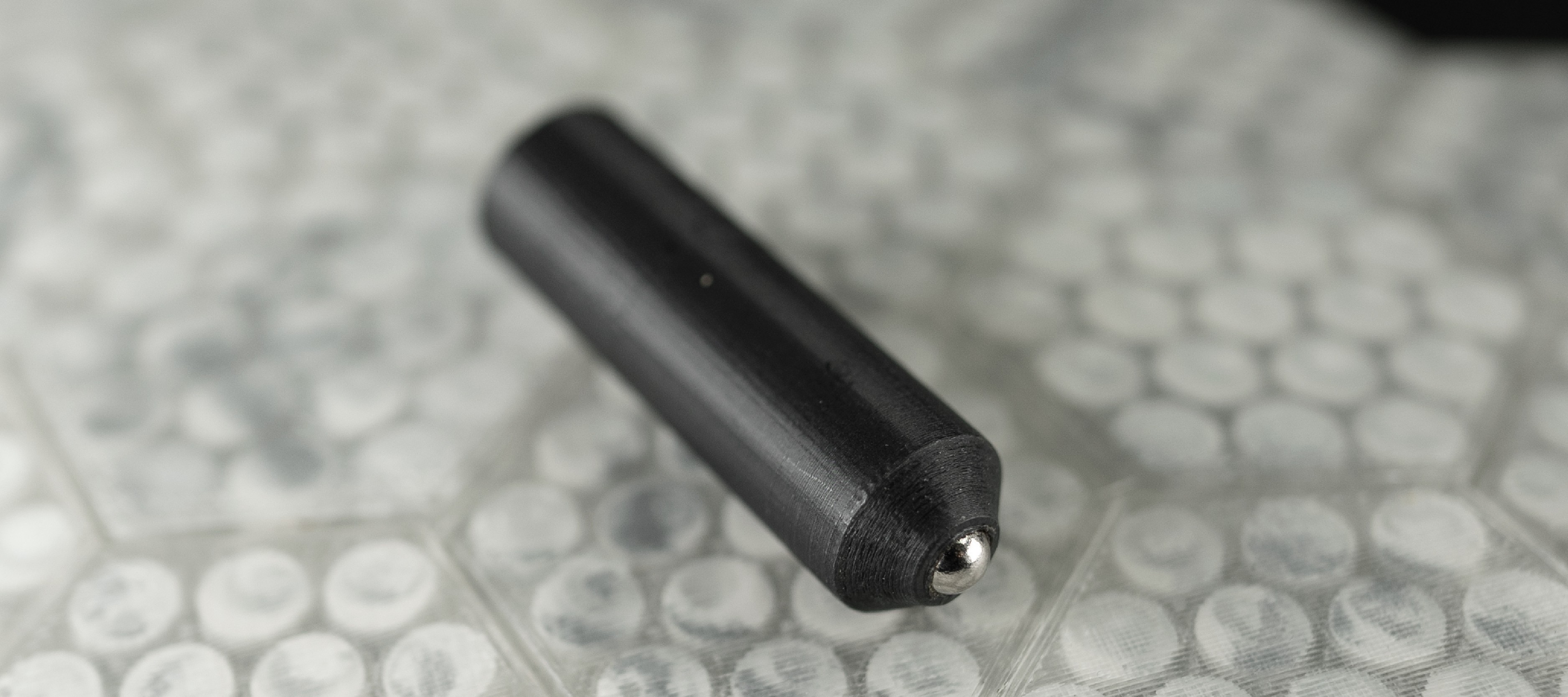}
  \caption{3D printed erasing pen with a magnet. We use this magnetic pen to edit the printed displays showcased in this paper.}
  \Description{Erasing Magnet Pen.}
  \label{fig:pen}
\end{figure}

As shown in Figure~\ref{fig:magDisplay}, a magnetophoretic display consists of cell matrices that are distributed across the display surface.
Each cell contains a mixture of opaque liquid and magnetic powder; the two components often have contrasting colors.
When a magnet (e.g., a magnetic pen as in Figure \ref{fig:pen}) approaches the surface of one or more cells, the magnetic powder of ``color A'' is attracted to the top, whereas the liquid of ``color B'' is pushed to the rear side of the cell, resulting in ``color A'' being presented at the surface.
The same principle applies when magnetic sources approach the back surface of the display, resulting in ``color B'' being displayed at the surface.
In addition, the combined liquid mixture needs to be viscous enough to prevent the powder from precipitating or freely moving within the cells for the display to be stable and always on.

While most magnetophoretic displays are in the form of 2D panels with the top cell made of a thin plastic membrane, we aim to bring the 2D display into the third dimension through 3D printing. 
Thus, our 3D editor should be able to convert an off-the-shelf 3D model into a shelled structure with distributed cells or voxels across its surface. 
Our hardware should be able to automatically print these cells watertight and in different orientations; their surface should also be thin enough for the powder inside to react to external magnetic forces on both its front and back side. 

In the following sections, we will first present our experiment on the material mixtures and the FDM 3D printer modification. We will then present our design editor and the modified G-code that parses a 3D model and automates the printing process.

\section{Fabrication of Mag display}

\subsection{Liquid Mixture} \label{injMatl}

The liquid mixture for magnetophoretic displays needs to meet two requirements. 
1) The magnetic powder should be easily movable from one side of the cells to the other, given the sizes of printable cells on a millimeter scale (See Section~\ref{printing parameters}). 
2) The liquid should be able to ``hold'' the powder in place when no external forces are applied. As the commercial magnetophoretic liquid used in Magna Doodles is not easily accessible, we proposed our own formula. 

\begin{figure}[h]
  
  \includegraphics[width=\columnwidth]{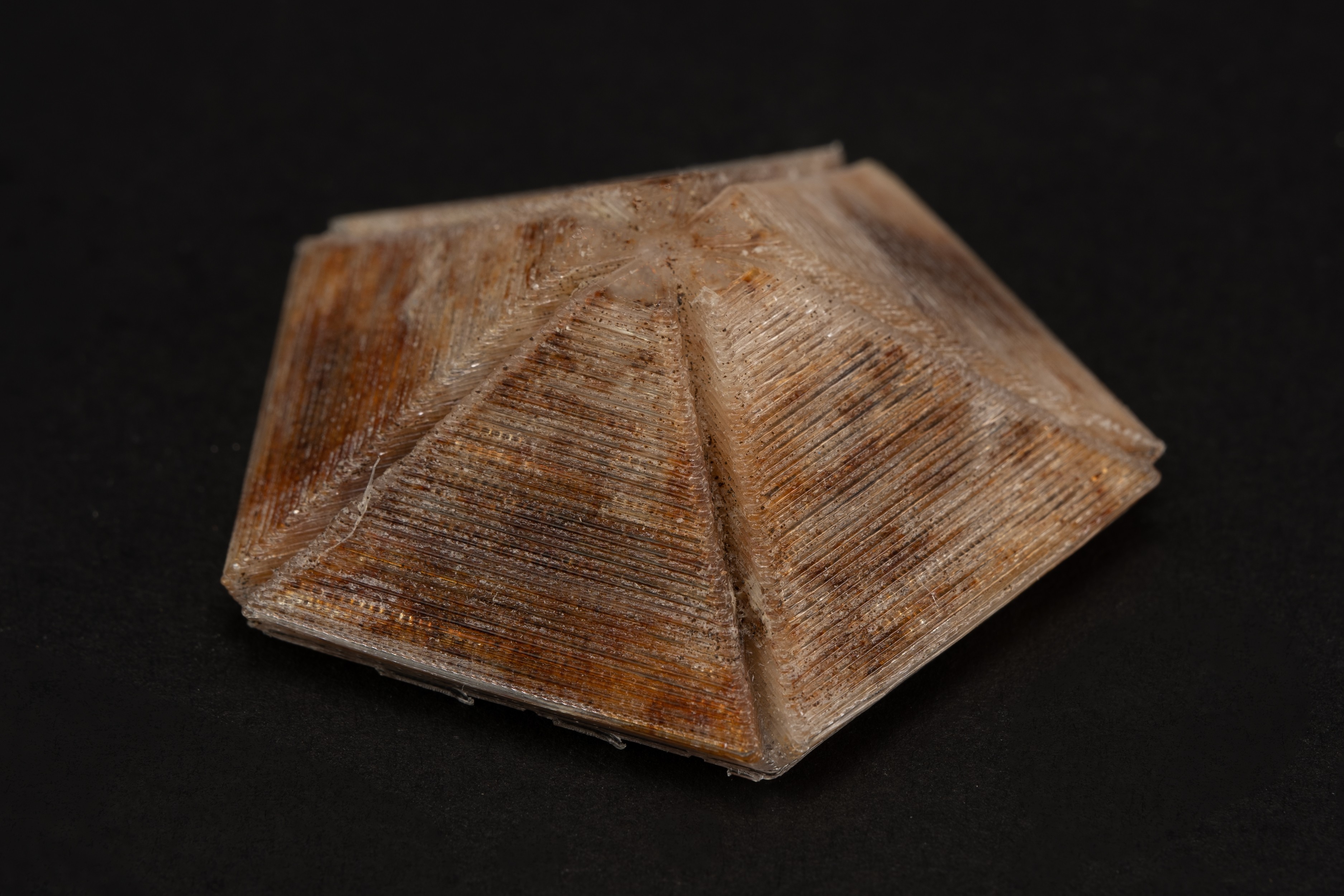}
  \caption{Iron powder rust when coupled with water based mixture.}
  \Description{Iron powder rust when coupled with water based mixture.}
  \label{fig:rust}
\end{figure}

To meet the first requirement, we adopt iron powder as our choice of magnetic powder due to its high permeability, superior magnetic susceptibility~\cite{magsus} and wide accessibility~\cite{ironPowder}.
For the liquid substrate, we experimented with both purified water and mineral oil by manually injecting them into printed test cells. The mineral oil is adopted as it prevents the iron powder from rusting in the wet environment over time (Figure~\ref{fig:rust}). The oil substrate is further thickened with talcum powder. 
Additional oil-based white dye is also added to the mix for better contrast.

\paragraph{Ratio.} The ratio of mineral oil, talc, iron powder, and dye in the mixture is critical for the visual effect.
The viscosity of the liquid increases monotonically as the talc-to-oil ratio increases, allowing the iron powder to remain in place when no external magnetic force is nearby. However, an excessive viscosity would also impede the movement of the iron powder, thereby reducing the responsiveness of the printed display.

Iron powder also has to be mixed in at a proper ratio so that it can completely cover the cell face when a magnet is present and leaves minimal residue when the magnet is removed. 

Following the rationale above, we empirically determine the mixture's ratio. 
For the examples presented in this paper, we adopt a weight ratio of 25:35:40:1 for mineral oil, talcum powder, iron powder, and coloring dye.

\subsection{3D Printer Modifications}\label{printer modification}
To automatically print 3D magnetophroetic displays, we added a custom duo-nozzle modification to an off-the-shelf, low-cost FDM desktop 3D printer (Creality Ender 3 Pro).

\begin{figure}[h]
  
  \includegraphics[width=\columnwidth]{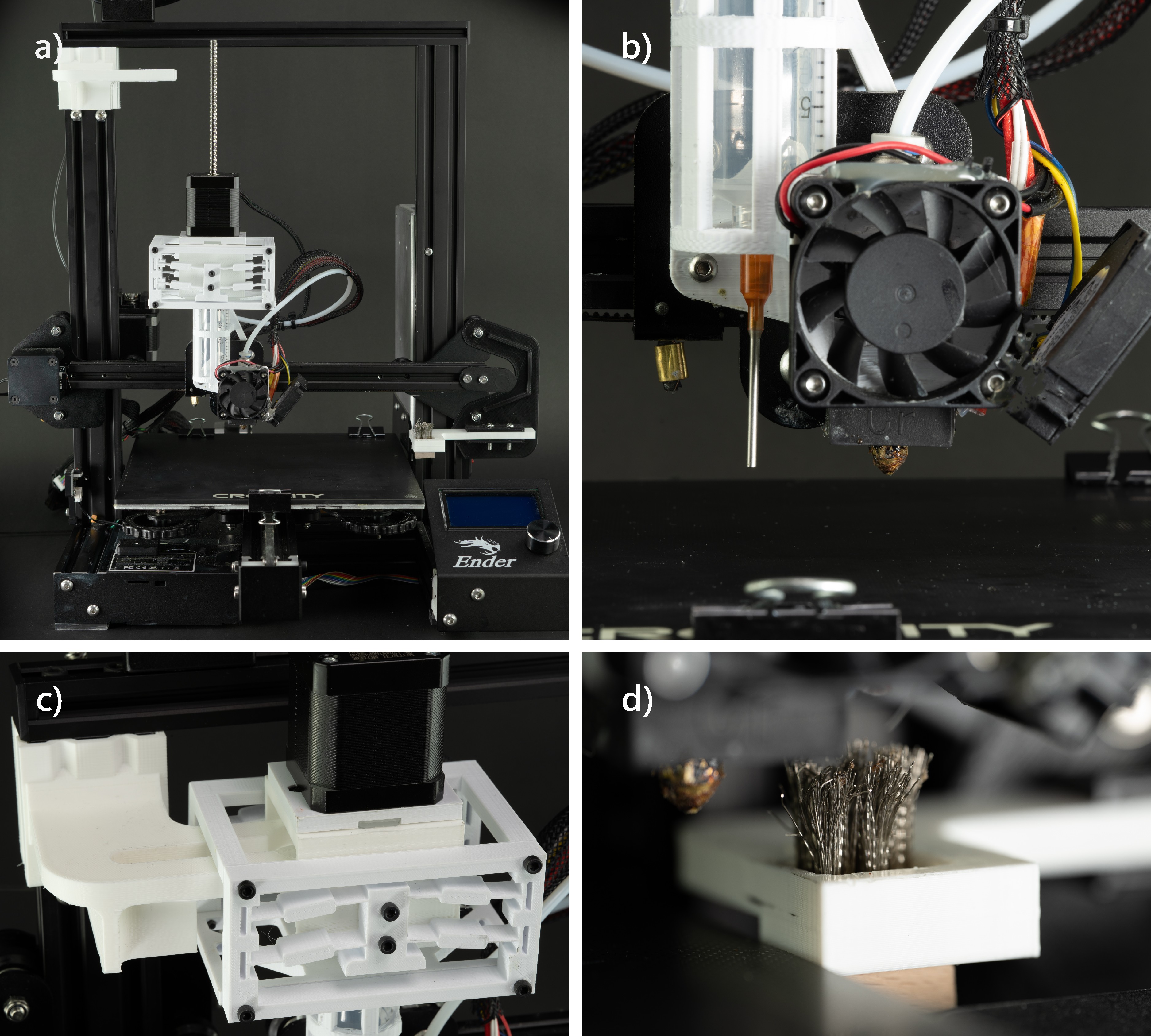}
  \caption{Modified Ender 3 Pro for magnetophoretic display printing. a) The printer with all the hardware modifications displayed holistically; b) the second stepper-driven injector for liquid mixture injection; c) the nozzle shifting mechanism that moves the injector up and down with the help of a stationary fork; d) a wire brush mounted for FDM nozzle cleaning.}
  \Description{Modified Ender 3 Pro for magnetophoretic display printing. a) The printer with all the hardware modifications displayed holistically; b) the second stepper-driven injector for liquid mixture injection; c) the nozzle shifting mechanism that moves the injector up and down with the help of a stationary fork; d) a wire brush mounted for FDM nozzle cleaning.}
  \label{fig:printer}
\end{figure}

Specifically, we added a stepper-driven syringe-based liquid injector next to the original FDM extruder for liquid injection (Figure~\ref{fig:printer}).
The liquid injector consists of a thread-shaft NEMA 17 stepper, a 30ml syringe with the corresponding piston, and a blunt needle.
To ensure consistent flow with the chosen liquid mixture ratio (See Section \ref{injMatl}), the blunt needle needs to have a minimum size of 14 Ga. Smaller needle outlets will frequently be clogged.
To avoid potential collisions, we also implemented a set of 3D printed bi-stable structures~\cite{bi-stable} that can keep the liquid injector up above or down below the FDM extruder during the printing (Figure~\ref{fig:printer}c). The shift of the liquid injector is activated by a 3D printed fork installed stationary on the printer's frame.
Finally, a wire brush is added to the far end of the X-axis of the printer to help clean and calibrate the FDM extruder after each liquid mixture injection (Figure~\ref{fig:printer}d).
The control of the injector is achieved using SKR 1.4 Turbo~\cite{skr1.4} and the Merlin 2.07 firmware~\cite{marlin}.

Although our hardware add-ons are designed for this machine, they can easily adapt to other Cartesian machines for bigger printing volumes.  

\subsection{Printing Routine}\label{printing parameters}
\begin{figure}[h]
  
  \includegraphics[width=\columnwidth]{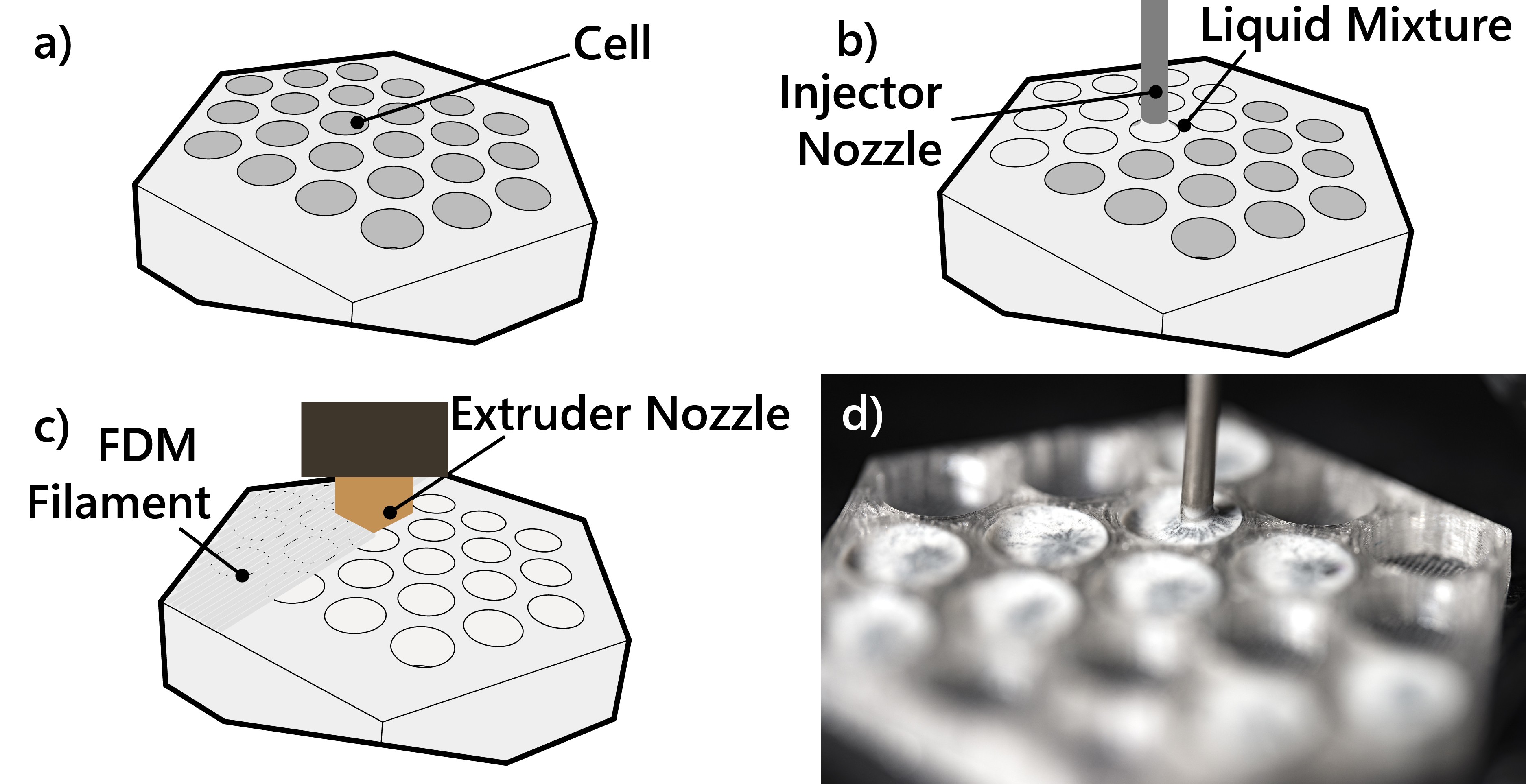}
  \caption{Illustrations of the printing process: a) cell ready for injection; b) liquid mixture injection; c) cell closed with FDM filament; d) the liquid injection in action.}
  \Description{Illustrations of the printing process: a) cell ready for injection; b) liquid mixture injection; c) cell closed with FDM filament; d) the liquid injection in action.}
  \label{fig:injection}
\end{figure}

The printing of magnetophroetic displays is similar to the conventional FDM process, where the printing materials are deposited layer by layer.
When the print proceeds, certain cell(s) would have the majority of their volume printed, leaving an opening that fits the liquid injector nozzle (Figure~\ref{fig:injection}a). 
The liquid injector will then be lowered with the fork and bi-stable mechanism and hover at the openings to inject the liquid mixture (Figure~\ref{fig:injection}b and d).
After injection for all required cells at this layer, the liquid injector is reverted back to a higher position, and the FDM nozzle resumes to print the next layer(s) and closes the cell(s) (Figure~\ref{fig:injection}c).

\subsubsection{Stable printing across layers} 
Since the injected liquid mixture is oil-based, oil that permeates the current layer of the print, if any, would affect the adhesion of the next layer. 
To accommodate this, a small, fixed-volume liquid retraction is applied after every injection to prevent excessive liquid from dripping into undesignated printing areas. 
While the retraction mechanism prevents failures caused by dripping, it may occasionally introduce air bubbles to the syringe nozzle. 
When air enters the nozzle, the retracted liquid volume is diminished, resulting in inaccurate injection of the liquid into the next cell(s). 
To ensure consistency across injections, an additional liquid dumping area is printed during the fabrication process. 
Every time a new series of liquid injections are needed, the liquid will first be ``over'' purged into the dumping area so that the liquid retraction will always start ``fresh''. 

Similarly, residue filament material may also drip from the FDM nozzle when the filament extrusion is idle for a considerable amount of time.
In our printing routine, we ``over'' retract the filament every time the printing process switches from liquid injection to filament extrusion, and purge extra filament from the FDM nozzle and wipe it on the wire brush.

With the above printing routine, we experienced very few failed prints because of the modifications. Most failures occur due to layer shifting or the first layer not sticking, which are common issues with consumer printers.

\subsection{Cell Parameters}\label{cell dimension}
To understand the feasibility and constraints of the printed cells, we experimented with a variety of cell properties, such as the printed cell's cross-section size, thickness, and depth. 
We also tested the overall display visibility. All the experimental models were printed with our modified 3D printer as described in Section~\ref{printer modification}.

\subsubsection{Cross-section size}
The cross-section of a cell is the size of each voxel when viewed from outside of the model. 
Its printability depends on the top opening of a cell at the time of liquid injection. 
As the opening is where the liquid is injected, it must have a minimum diameter greater than that of the liquid injector nozzle to ensure the liquid mixture is injected only within the target cell and does not overflow into its neighbors. 
With our current setup, the nozzle diameter is 14 Ga, or 2.1 mm. 

\begin{figure}[h]
  
  \includegraphics[width=\columnwidth]{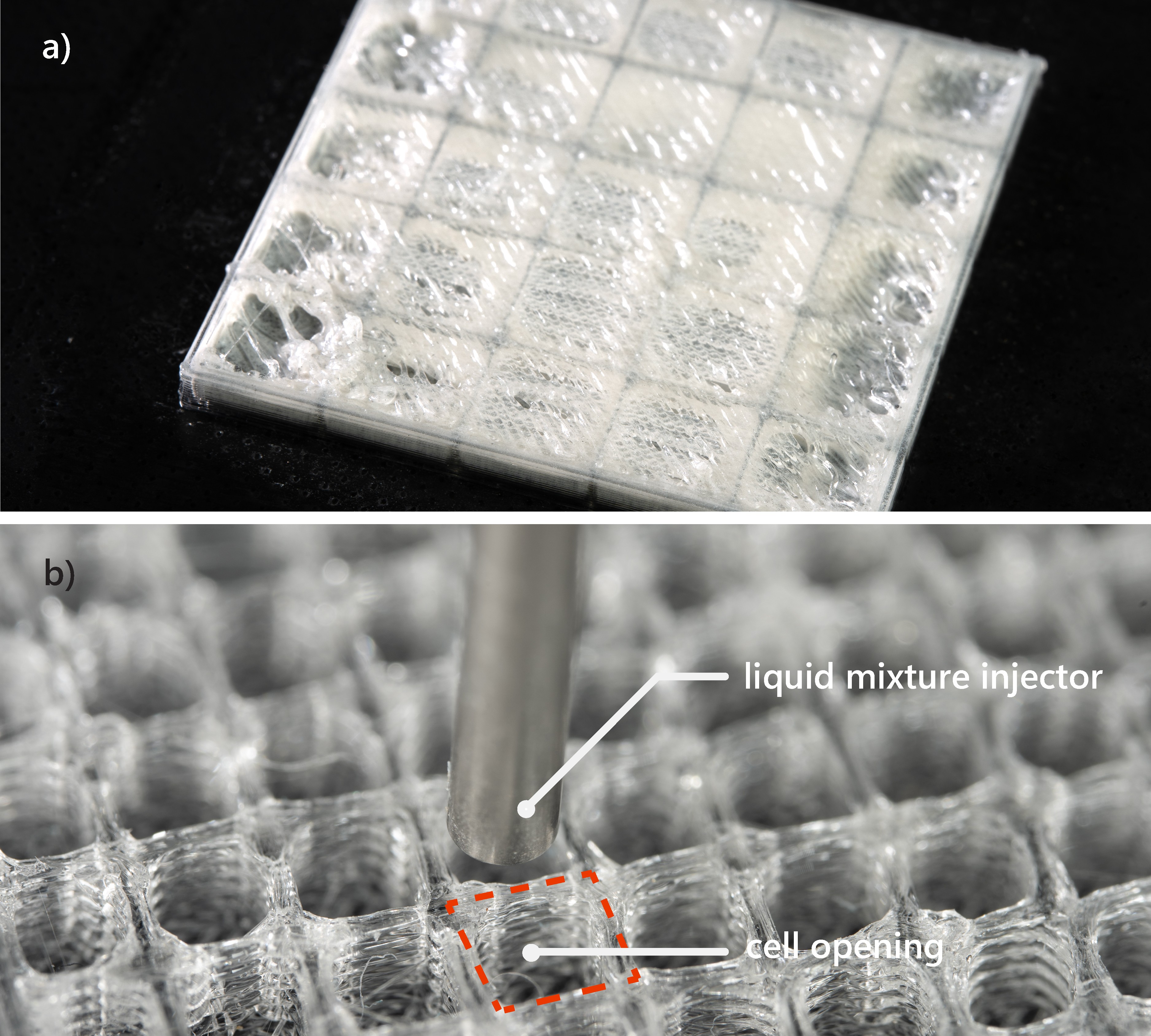}
  \caption{Cell size constraints: a) cell size smaller than the injector nozzle diameter; b) cell size too large to close with next layers' of filament.}
  \Description{Cell size constraints: a) cell size smaller than the injector nozzle diameter; b) cell size too large to close with next layers' of filament.}
  \label{fig:cellSize}
\end{figure}

The opening cannot be infinitely large either, as it must ensure the closing layers of the cells, which are predominantly overhanging structures, can be successfully printed. We report that with our current printer setup, the reliably printed overhang distance is 7 mm.
Thus, the printable cell size will have an inscribed sphere diameter ranging from 2.5 mm to 6.5 mm (Figure~\ref{fig:cellSize}).

\subsubsection{Surface thickness}
The front and back sides of the cell are critical to the success of the printed display, as the front side displays the visual information while the back side activates the magnetic powder, i.e., updates the display visuals. 
We have experimented with printing the surface with one to five layers of transparent filament. 
We empirically decided on the printing with the two to three layers of filament with a thickness of 0.6 to 1 mm, as the printed surface can reliably seal the liquid while being a clear window to reveal the iron powder's black color.

\subsubsection{Depth}
The depth of a cell is the distance between its front and back surfaces, as well as the distance that iron powder travels. 
The main deciding factor for the depth is to ensure the magnetic powder can be fully attracted to one side of the surface. 
For all models printed in this paper except those mentioned specifically, the cell depth has a maximum of 5 mm.

\subsubsection{Overall visibility}
In previous sections, we examined the printability and the size constraints of a single cell. 
Here we briefly discuss the resolution of a matrix of cells and how the cross-section cell sizes may affect the overall visibility.

\begin{figure}[h]
  \includegraphics[width=\columnwidth]{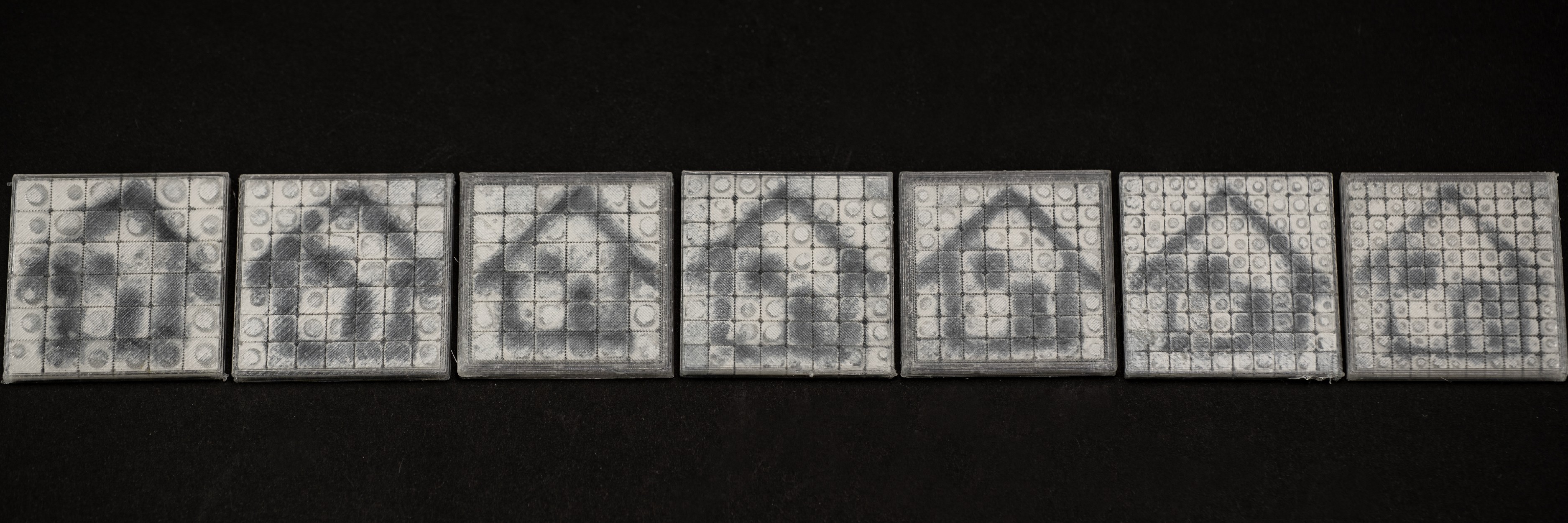}
  \caption{Seven test printing samples. The stroke width and the details of the drawing are affected by the cell size. }
  \Description{Seven test printing samples. The stroke width and the details of the drawing are affected by the cell size.}
  \label{fig:visibility}
\end{figure}

In general, we can reliably print 3D models whose cell parameters adhere to the aforementioned cell constraints, with a higher density of cells per unit area resulting in better and more detailed visual effects. 
For example, Figure~\ref{fig:visibility} presents 7 test printing samples, each with a 40 mm by 40 mm display area and a cell edge length that decreases from 6 mm to 3 mm in increments of 0.5 mm. 
While all samples are able to demonstrate the overall shape of the house icon, the samples with smaller cell size shows higher rendering resolution in stroke width, and only the last one is able to show the door of the drawing clearly.

\section{Design Editor}\label{design tool}

\begin{figure}[h]
  
  \includegraphics[width=\columnwidth]{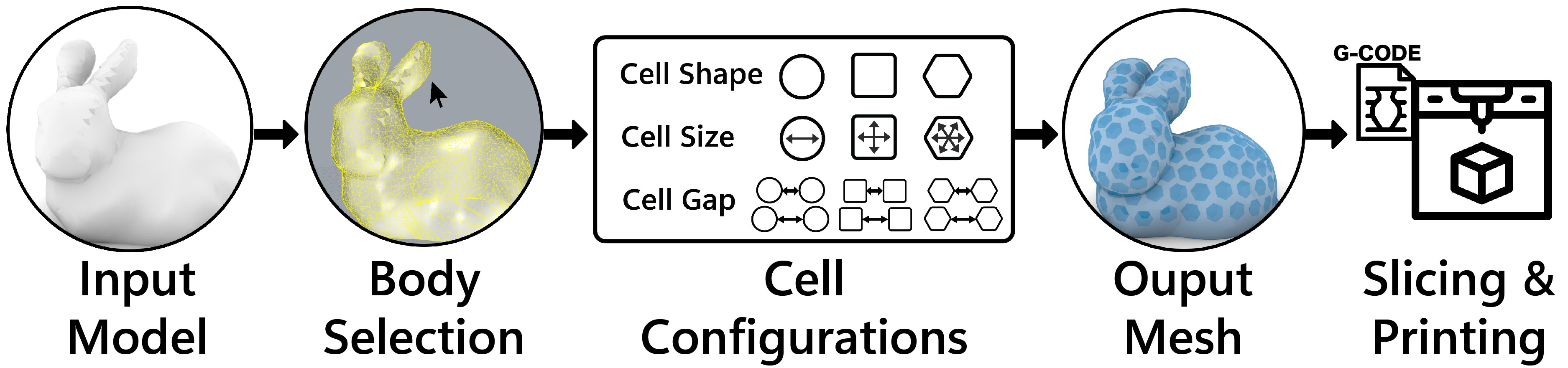}
  \caption{The workflow of using the design tool to create a cell-based 3D model for slicing and printing. }
  \Description{This figure shows the workflow of using the tool to generate a cell-based model for fabrication.}
  \label{fig:tool}
\end{figure}

We develop a Rhino 3D plugin that allows the end-user to convert a 3D model into a magnetophoretic display (Figure \ref{fig:tool}).

End users will first configure the shapes (i.e., circle, square, or regular hexagon), sizes, and the gaps between cells to specify how the converted model will look. 
A live preview will be generated to help end users visualize and validate the conversion. 
Once the configurations are confirmed, a series of 3D cells will be automatically generated and embedded in a shell-like model. 
All the cells are sandwiched between the body's exterior and interior surfaces with sufficient room for the liquid mixture to flow. 
The cell-based model can be exported to a custom slicer, which converts generated meshes into G-code for fabrication. 

The design editor is developed in C\# with Rhinocommon API \footnote{Rhinocommon API: https://developer.rhino3d.com/} and Human UI \footnote{Human UI: https://grasshopperdocs.com/addons/human-ui.html}, a Rhino Grasshopper add-on. 
The custom G-code post-processor is developed in Python with Trimesh \footnote{Trimesh: https://trimsh.org/index.html}, a triangular mesh processing library. Below, we describe the user interface and the computational procedure for converting a 3D mesh into a cell-based shell in the design editor.

\subsection{User Interface}

\begin{figure}[h]
  
  \includegraphics[width=\columnwidth]{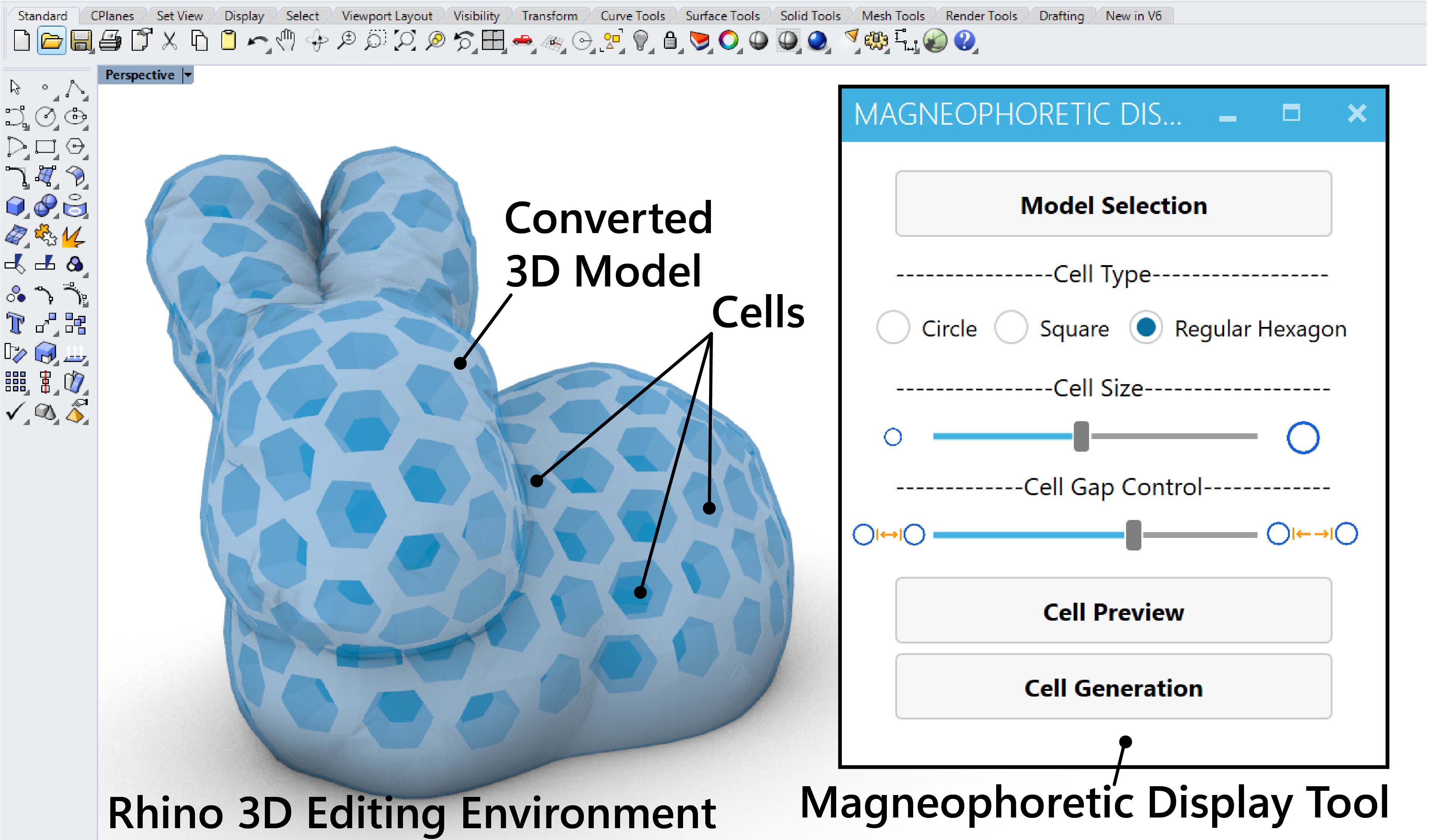}
  \caption{The user interface of the design editor. }
  \Description{The user interface of the design editor.}
  \label{fig:interface}
\end{figure}

The user interface has four functions (Figure \ref{fig:interface}): body selection, cell configurations, cell preview, and model generation. The corresponding UI components are arranged from top to bottom accordingly so that the 4-step workflow is easy to follow.

\subsubsection{Body selection} As the first step, the user chooses the 3D model in the Rhino editing environment with the \textit{Model Selection} button. 
The selected model is then converted into an initial polygonal mesh and highlighted in the scene.

\subsubsection{Cell configurations} The user can select the cell's cross-section shape with three options: circle, square, and regular hexagonal cells. For the selected cell type, the user can parameterize a number of settings by dragging the individual sliding bars in the interface. For example, the user is able to adjust the circular cell's diameter, the square cell's side length, and the diagonal length of the regular hexagonal cell. 
They can also adjust the distance between adjacent cells. For each adjustment, we implemented the upper and lower bounds based on the empirical data reported in Section~\ref{cell dimension}.

\subsubsection{Cell preview} After completing the initial configuration of the cells, the user can validate the design by clicking on the \textit{Cell Preview} button. A visual rendering will be updated in the Rhino editing environment, illustrating the distribution of all cells across the surface of the model, which may help the user adjust their design if necessary.

\subsubsection{Model generation} As the final step, the user can click the \textit{Cell Generation} button and create the final model. The original mesh will be first converted into a shell structure. All cells will then be generated within the shell layer, with top and bottom screen layers measuring \SI{0.6}{\milli\meter}.

\subsection{Cell-Base 3D Model Generation} \label{cellGen}
We now introduce the computational procedure for converting a 3D model into a shell-like mesh with distributed watertight cells (illustrated in Figure~\ref{fig:algorithm}). The procedure has two steps: 1) generating the shell-like mesh model; 2) generating cells and combining them with the shell-like mesh model.

\subsubsection{Generating the shell-like mesh model}
\begin{figure}[h]
  
  \includegraphics[width=\columnwidth]{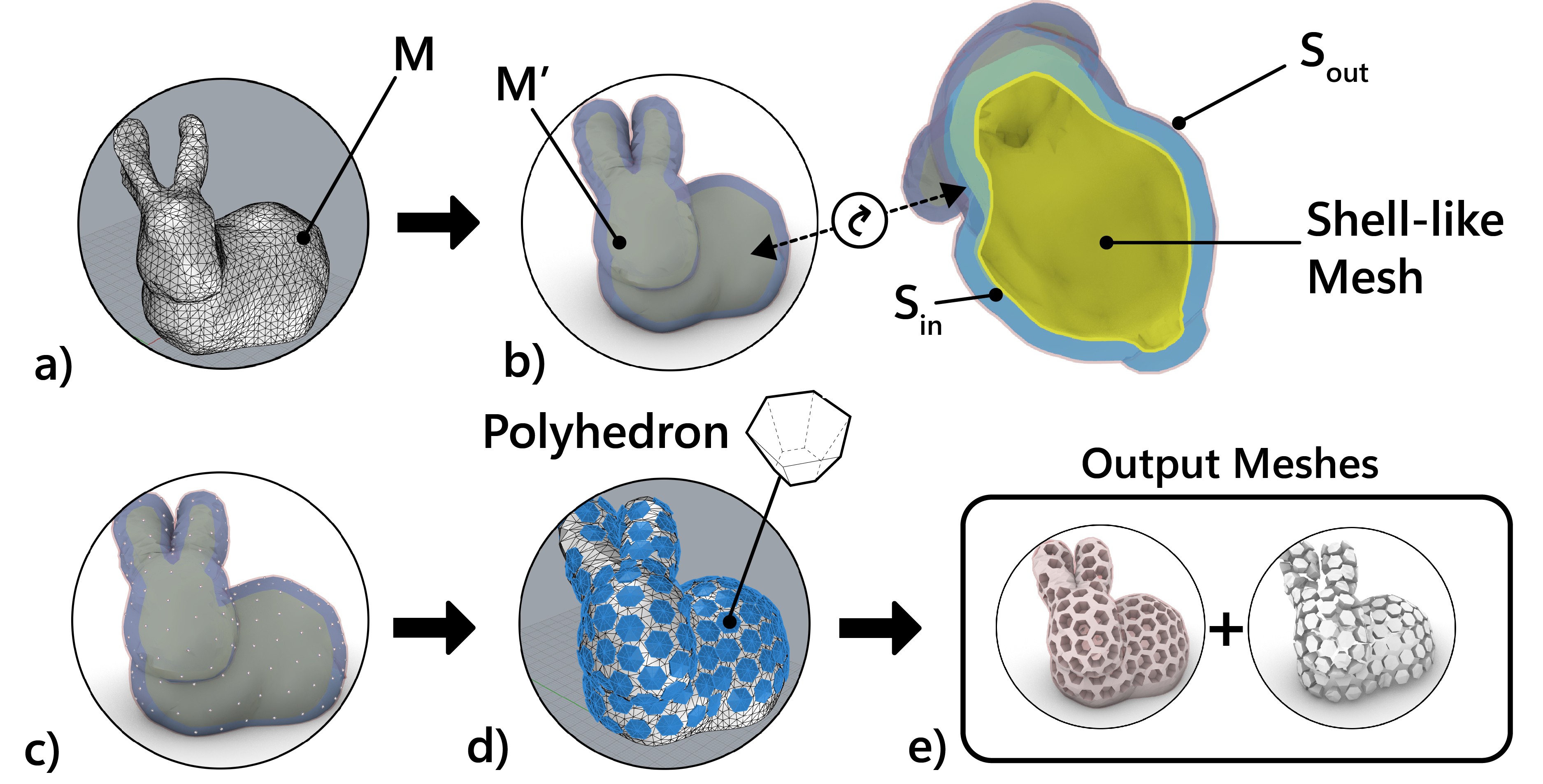}
  \caption{The process for generating cells in the shell-like output mesh.}
  \Description{The cell generation process}
  \label{fig:algorithm}
\end{figure}

To create a display that allows for both drawing and erasing, a 3D model needed to be converted into a shell-like mesh so that a magnet could reach the surfaces from both the outside and inside of the model. To create the shell, our editor first converts the selected model into a polygonal mesh $M$ (Figure~\ref{fig:algorithm}a), then generates the outer ($S_{out}$) and inner ($S_{in}$) shell layers that serve as the mesh model surfaces (Figure \ref{fig:algorithm}b). Specifically, to generate $S_{out}$, we first offset $M$ by the distance of $H_{cell}$ (the cell height) to create a new larger mesh $M'$, and then offset $M'$ by another $H_{os}$ (the screen thickness) to create a new mesh as the outmost surfaces. The volume between the newly generated mesh and $M'$ is retained as the outer shell layer $S_{out}$. Similarly, $S_{in}$ is generated by first creating a model that is offset from $M$ inward by $H_{os}$, and then retaining the volume in between. The offset operations are based on \verb|Rhinocommon OffsetMesh| command. The intermediate mesh $M'$ is used for cell generation in the next step.

\subsubsection{Generating cells}\label{cell generation}

The cells to be generated will be sandwiched between $M$ and $M'$ and are thus adjacent to the two shell layers $S_{out}$ and $S_{in}$. The cell generation can be further described as a two sub-step procedure: 1) retopologize $M'$ and 2) cell polyhedrons computation.  

The reason that we need to re-mesh $M'$ is to convert it into an isotropic triangular mesh, where the triangles are equally distributed across the surface so that their corners can be used as the cells' projected center locations. The re-meshing is done by calling Instant Meshes processor~\cite{jakob2015instantmeshes} as an automatic background processing. The number of the triangles in the re-meshed $M'$ is calculated based on the user's input on the cell size and the in-between gap.

After remeshing, each of the triangle corners of $M'$ corresponds to the cell's location. Based on the user's selection of the cell's cross-section shape, we generate the corresponding convex polyhedron, i.e., truncated cone (circle), truncated pyramid (square), and truncated hexagonal pyramid (regular hexagon). 
Each of the convex polyhedron is computed with perspective transformation where the cross-section cell is projected to the bottom surface of $S_{in}$ and the top surface of $S_{out}$. The loft between the two projections is the resulting polyhedron at the target cell's location, which is created using the \verb|Rhinocommon CreateSoild| command.  

After all the convex polyhedrons are generated, they are  used to create a porous mesh body (see the gray cells in Figure \ref{fig:algorithm}e). The top shell layer, the bottom shell layer, and the porous mesh body, combined, compose the final mesh model that is ready for slicing. All generated cells are also exported as individual meshes, which is used in the slicing procedure below.

\subsection{Slicing}

To automate the printing process with our modified duo-nozzle printer, the machine must inject the liquid mixture at the correct height for each cell. 
To accomplish this, we first use the commercial 3D printing software Cura to parse the model and generate the initial G-code. 
The injection heights and locations are then computed, and the G-code is modified accordingly with a python script.

The initial G-code is generated using Cura's default settings, with the exception that the printing speed is reduced to between 50 to 60 percent of the original. ``Enable bridge settings'' is also checked with the default parameter.
The support structure is only added to external overhanging structures, leaving the cells empty.
After the initial G-code is created, we use the cell meshes generated in Section~\ref{cell generation} to compute the injection location for each of the cell, and adjust the G-code accordingly. 

\subsubsection{Finding injection location for a cell} 
\begin{figure}[h]
  \includegraphics[width=\columnwidth]{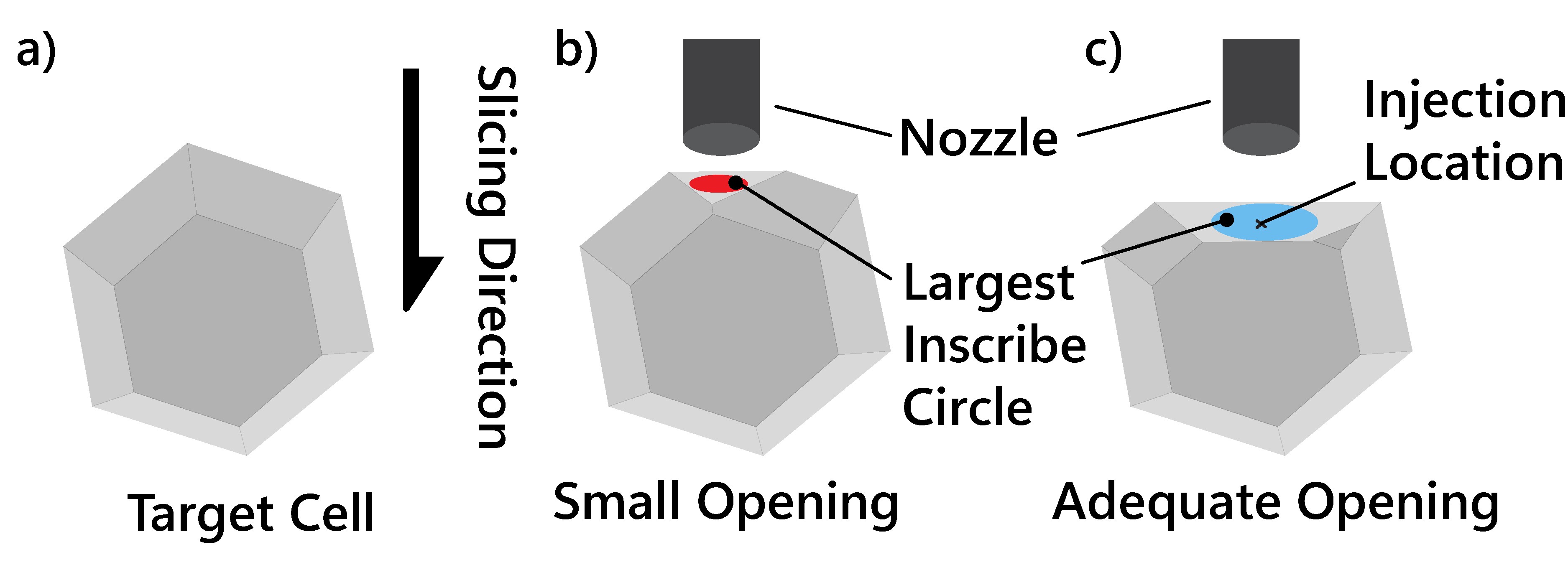}
  \caption{Finding injection location from top to bottom at each layer height.}
  \Description{Finding injection location from top to bottom at each layer height.}
  \label{fig:gcode}
\end{figure}

Each cell's mesh is sliced with every FDM layer thickness in height starting from the top.
At each layer height, the intersection polygon of the cell mesh and the XY plane at this given height can produce the largest inscribed circle and its center location (Figure \ref{fig:gcode}b). 
When the diameter of the inscribed circle is larger than the nozzle diameter used, we take the current layer height and the central location of the inscribed circle to form the injection location X, Y, and Z for this cell (Figure \ref{fig:gcode}c). 
Otherwise, the slicing of this cell will keep going until the lower part of the cell has a volume less than 80\% of the entire cell volume. 

\subsubsection{Generate G-code for each injection layer}
Once the injection locations are found, they are sorted according to the heights of injections.
G-code blocks for each injection are generated according to the location and the partial volume of the cell below its injection height.
Since the heights are parsed with the same layer thickness used for the FDM slicing, all injections' G-code can be appended after the end of a corresponding layer of FDM deposition.

\subsubsection{Purging materials when switching printing methods} 
As described in Section~\ref{printing parameters}, both the filament and liquid mixture need to be purged and recalibrated for a stable printing process. Thus in addition to calculating the injection locations, we insert a fixed G-code module whenever the printing switches between extrusion and injection. The fixed G-code module will direct the injector to eject the liquid mixture into the ``dumping area'' and to wipe the filament extruder. 

With the modified G-code, the entire printing process is automatic. All examples in the following section are printed using this process.

\section{Application}
In this section, we highlight a few examples to demonstrate the capabilities and opportunities that our magnetophoretic display printing pipeline enables.

\begin{figure}[h]
  \includegraphics[width=\columnwidth]{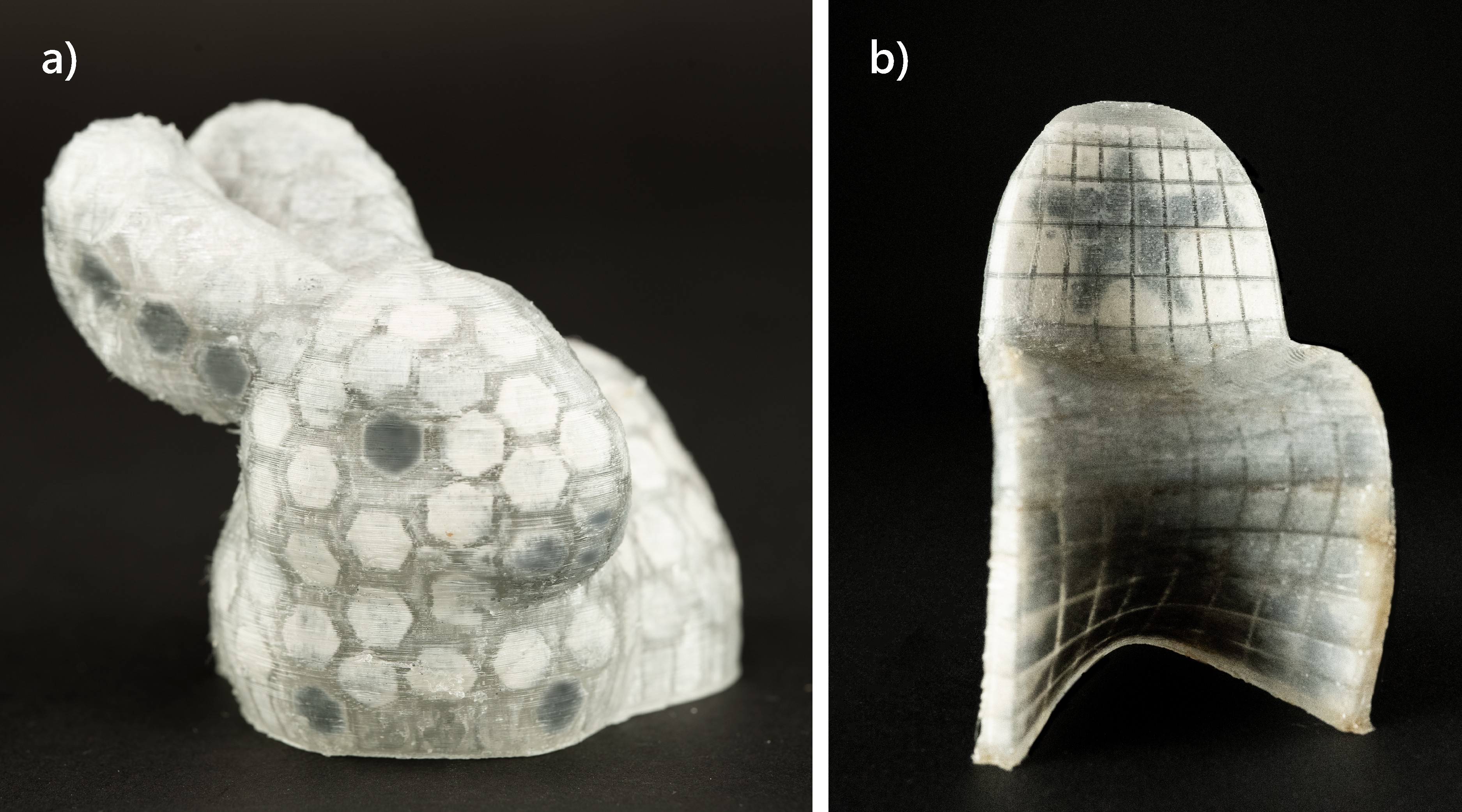}
  \caption{a) Stanford bunny model: The eyes, legs, and ears are post-edited; b) Panton chair.}
  \Description{a) Stanford bunny model: The eyes, legs, and ears are post-edited; b) Panton Chair.}
  \label{fig:bunny}
\end{figure}
\subsection{Stanford Bunny and Panton Chair}
The main promise of our proposed pipeline is to convert a 3D artifact into a magnetophoretic display whose appearance can be post-edited. Here, we showcase a Standford bunny model being printed using our pipeline.
With our 3D editor, we first convert the surface of the Stanford bunny into patterned hexagon voxels with a 4 mm distance across the corners. The model's abdomen is then hollowed automatically with the editor. With the custom G-code, the model is printed one-shot with our modified printer. Figure~\ref{fig:bunny}a presents a bunny model with post-decorated eyes, legs and ears. %
In addition, we also printed a down-scaled Panton chair to showcase a 3D display on a non-developable surface that can be viewed and operated on both sides (Figure \ref{fig:bunny}b).

\subsection{Espresso Cup as Post-it}
\begin{figure}[h]
  \includegraphics[width=\columnwidth]{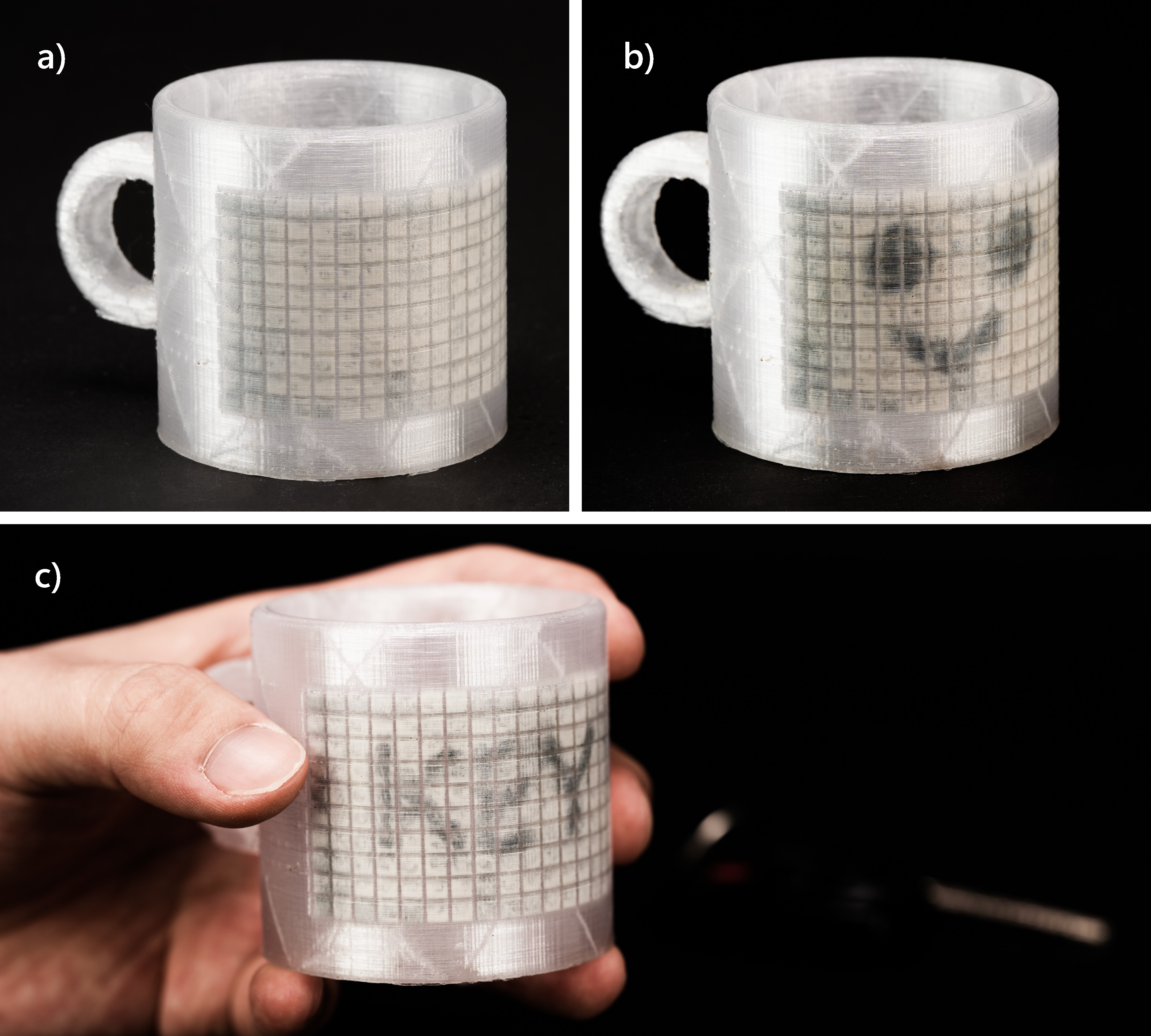}\caption{Espresso cup as post-it.}
  \Description{Espresso cup as post-it.}
  \label{fig:mug}
\end{figure}

Post-it or sticky notes are commonly used to leave messages and reminders for ourselves and others. Sticky notes can only adhere to flat surfaces, limiting their usefulness. In this example, a round espresso cup is converted into a post-it cup. As illustrated in Figure~\ref{fig:mug}, the front of the cup is converted and printed with 16 by 10 square cells, making the curved surface a drawing area. Figure~\ref{fig:mug}b and c illustrate two examples of notes. 

\subsection{Personalized Clothing}
Clothing is a form of self-expression for many people. Personalized designs have been used to make one-of-a-kind clothing, such as T-shirts with custom slogans or logos. However, the majority of wearables will retain the same appearance and do not have the capability for re-patterning.

\begin{figure}[h]
  \includegraphics[width=\columnwidth]{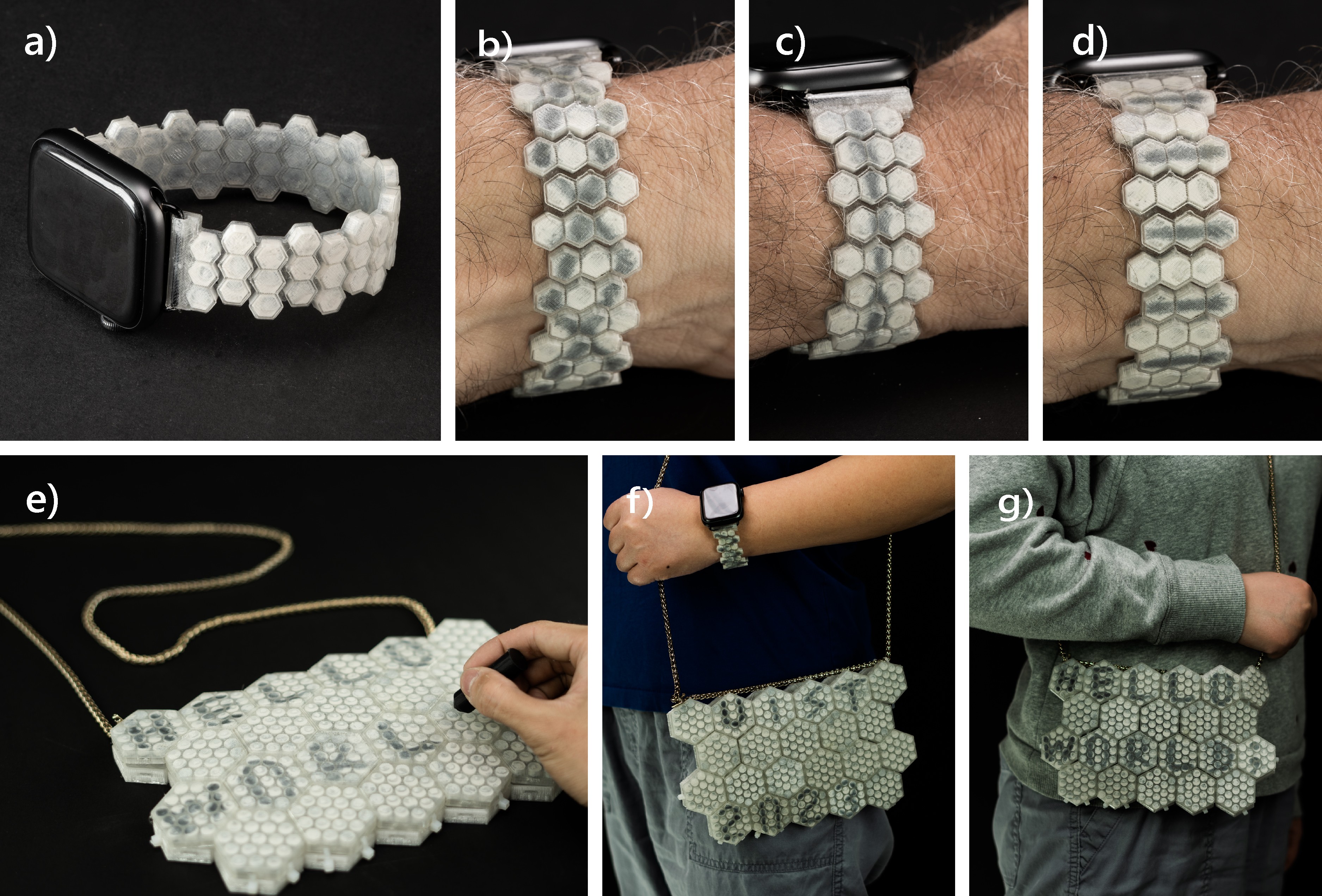}
  \caption{Wearable accessories. a-d) flexible watch strap with custom decoration. e-g) Tiled handbag with personal slogans.}
  \Description{Wearable accessories. a-d) flexible watch strap with custom decoration. e-g) Tiled handbag with personal slogans.}
  \label{fig:clothing}
\end{figure}

Here, we showcase two wearable accessories—one smartwatch strap and one handbag—printed using our pipeline. As shown in Figure~\ref{fig:clothing}a-d, the watch strap is printed with flexible TPU filament so that it can be bent and fit user's wrist. 
The decorative pattern can be easily changed across the surface. Similarly, Figure~\ref{fig:clothing}e-g showcase a handbag assembled with a handful of hexagon tiles that become magnetophoretic displays composed of circular cells. The handbag can be re-patterned with different slogans as a new look to match different fittings.

\subsection{Board Game Figurine}\label{miniatures}
\begin{figure}[h]
  \includegraphics[width=\columnwidth]{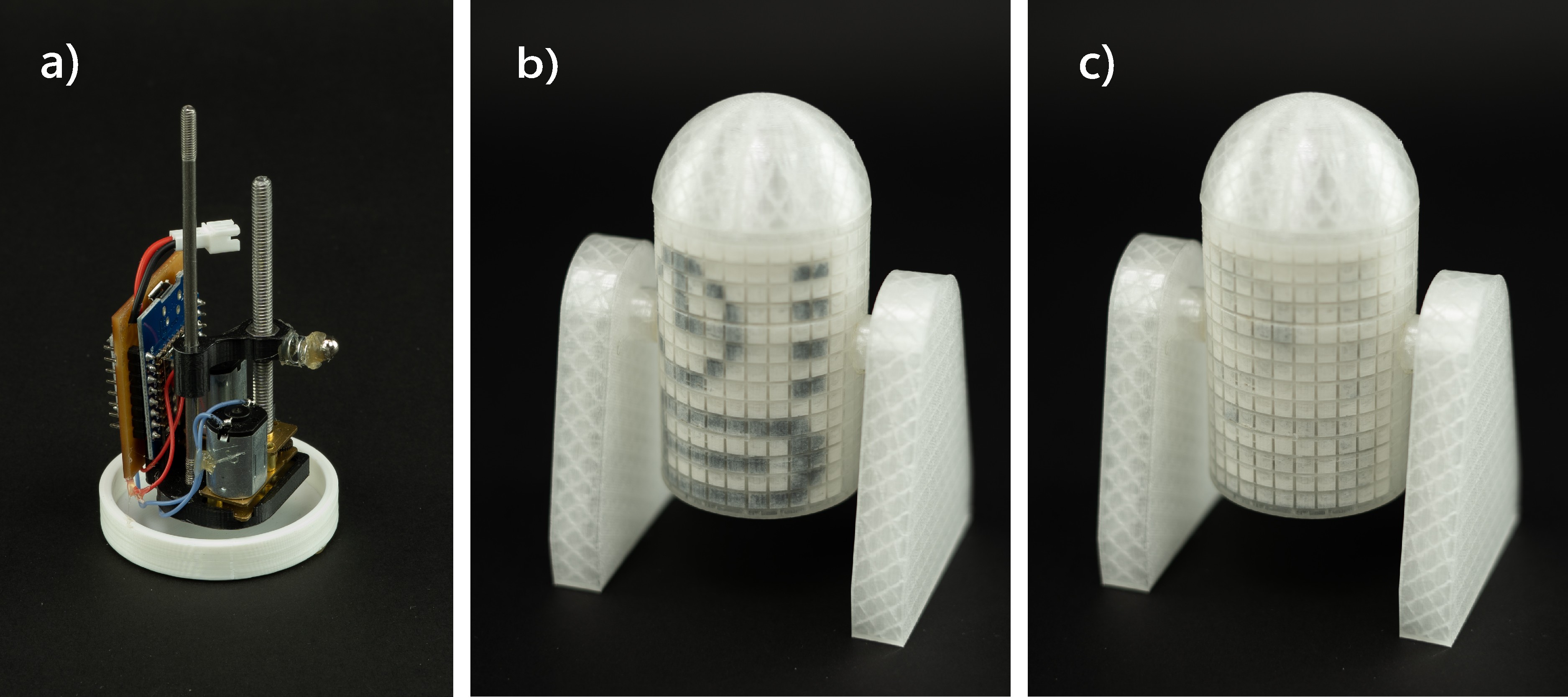}
  \caption{Board game figurine. a) a wireless motorized platform with an magnetic end-effector. It can be used to update the figurine in b) and c) digitally.}
  \Description{Board Game Figurine.}
  \label{fig:boardgame}
\end{figure}

Tangible figurines can enhance board game experience with embodiment. As most miniatures are static objects, dynamic gaming features (e.g. armor upgrade or downgrade) cannot be applied directly to a game figurine by automatically erasing existing patterns drawn on the figurine. 
In this example, we demonstrate that a 3D printed miniature can automatically update its appearance based on the progression of a board game by erasing existing patterns drawn on the figurine. 
In Figures~\ref{fig:boardgame}b, and c, certain patterns of a figurine are computationally removed, denoting various feature updates throughout a board game. 
The erasure of these patterns is achieved using a wireless motorized platform, as shown in Figure~\ref{fig:boardgame}a. Its magnetically equipped end-effector can move along a cylindrical surface, removing already-drawn patterns.

\section{Discussion and Future Work}
\subsection{Display Limitations}
While our pipeline enables the first 3D printable custom display that is self-contained, always-on, and interactive, the use of magnetic liquid as the display mechanism has several trade-offs.

\begin{figure}[h]
  \includegraphics[width=\columnwidth]{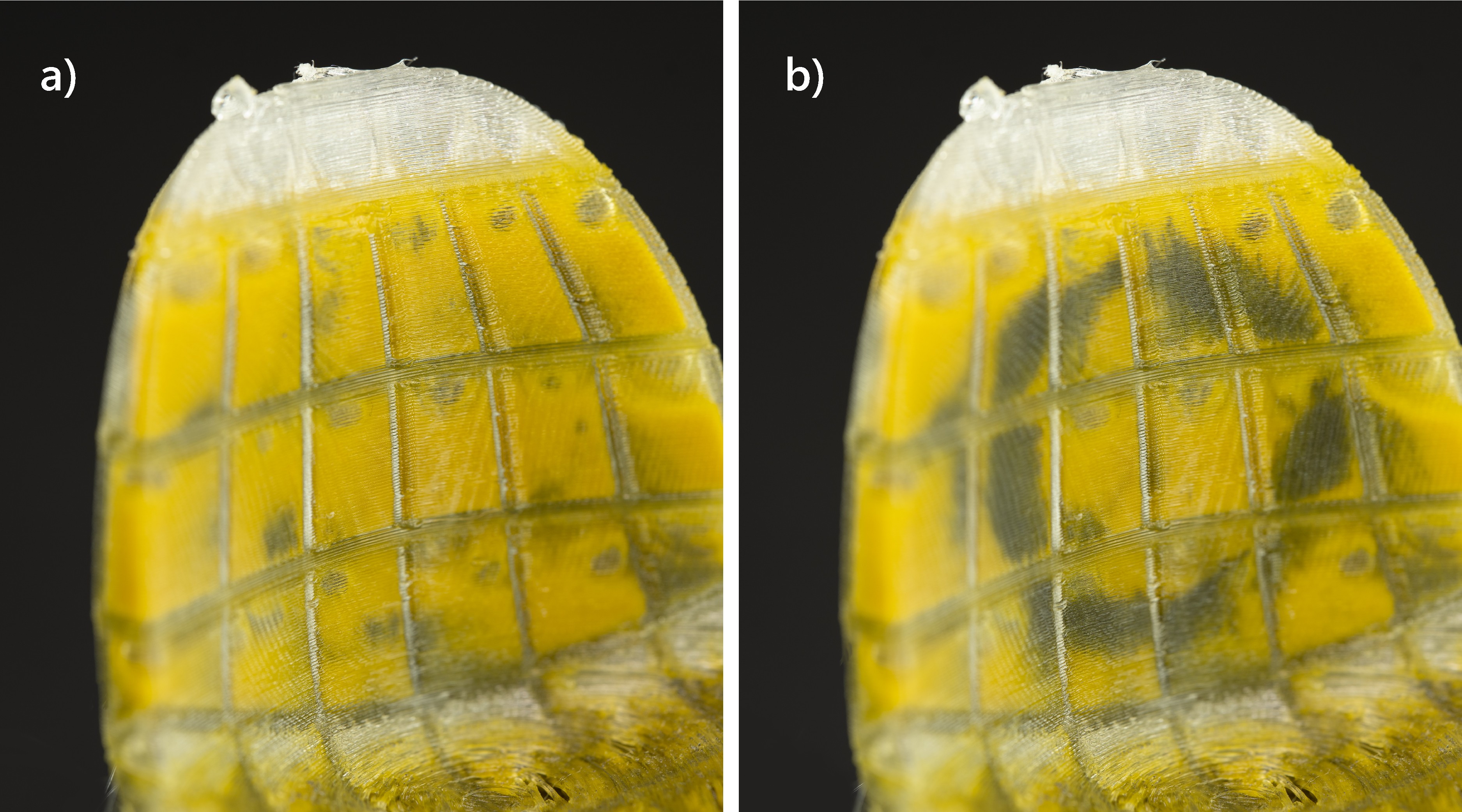}
  \caption{Printed sample with yellow oil-based dye.}
  \Description{Printed sample with yellow oil-based dye.}
  \label{fig:color}
\end{figure}

\subsubsection{Color composition} 
One of the limitations is that the liquid mixture can only display two contrasting colors: the black color from the iron powder and the white color based on the oil-based dye. To expand the color palette, we have explored different oil-based color dyes, such as the yellow dye in Figure~\ref{fig:color} as a white dye replacement. This can expand the display's color scheme beyond black and white.
We may further diversify the color scheme by substituting the black iron powder with other powder materials, such as those coated with oxide colorants~\cite{colorPowder}. While the working principle of the display suggests that the color palette will be limited to two hues, future machines might also use multiple injector systems, each of which is equipped with a different set of colored liquids that can be blended in-cell to create a wider range of colors.

\subsubsection{Stability} As the printed displays are activated with magnets, their appearances are susceptible to strong, external magnetic forces. 
While this is impossible to mitigate entirely, we provide anecdotal evidence that when used in daily life, the display patterns are fairly stable. 
For example, we have worn a smart watch with the patterned wrist strap (Figure~\ref{fig:clothing}b) for three days straight while performing daily activities, and have observed no pattern decay. 
This is most likely due to the proportional, inverse relationship between magnetic force and the distance from the surface. 
A N42 neodymium magnet that is placed 5mm away will have no impact on the display pattern.

Another concern is for how long the magnetic liquid mixture will remain active after the prints, or whether the printed magnetophoretic artifacts will be suitable only as temporary displays. To understand this, we printed a matrix of 16 by 10 with square cells with a 3 mm side length. We stored the sample at room temperature for over 120 days. Over time, we have observed no agglomeration or sediment of the iron powder mixture, and the display sample has remained in the same working condition.

\subsubsection{Digitally control the display}
One last limitation of our proposed method is that the patterns of the display are hard to control digitally. In Section~\ref{miniatures} we showcased a custom-designed actuator system that can update the display figurine computationally. However, we acknowledge that such an actuator requires custom design and is not easily generalizable, especially if the three-dimensional display has a complex surface geometry. 

While this is the limitation of magnetophoretic displays, our pipeline can be easily extended with different powder compositions to selectively control each of the cell displays. For example, FabricatINK~\cite{hanton2022fabricatink} demonstrated that E-Ink from upcycled E-Readers can be used for bespoke displays. Such particles can likely replace the iron powder in the current mixture. Combining with 3D-printable electrodes or custom flexible circuits \cite{10.1145/3526113.3545652}, we could potentially print digitally controllable 3D displays using the same pipeline proposed in this paper.

\subsubsection{Erasure method}
The erasure of the magnetic powder presented in this paper is done by applying magnetic force at the ``back'' of the display as described in section 3 \ref{principle}. 
Specifically, we place a magnet at the end of a stick that can reach the inside of each model, including the bunny ear, where the chamber has sufficient inner space. 
When the space at the back of any cell is smaller than the size of a magnet (currently 3mm in diameter), the cells cannot be erased from the inside.

Inspired by the shake-to-erase feature of toys like ``Etch A Sketch'', we have experimented with the ratio of an injected mixture with lower viscosity, hoping to allow the injected powder to redistribute upon shaking. 
We report that a viscosity lower than the mixture ratio of 20:20:40:1, as detailed in section \ref{injMatl}, will enable the shake-to-erase feature.
However, the powder would precipitate in less than 5 minutes and the display couldn’t hold its information for longer. 
The design decision should be made upon whether to prioritize the preservation of the display information or to enable the shake-to-erase feature.

\subsection{Tool Improvement}

While our design tool can convert a 3D model into a magnetophoretic display, we enumerate the challenges we encountered during cell generation and discuss potential future expansions of the tool. 

\subsubsection{Overlapped cells}
One limitation of our current tool is that overlapped cells may be generated at locations with very large curvatures. This is because the tool forms cells by lofting two polygons that are projected on $S_{in}$ and $S_{out}$ (Figure~\ref{fig:tool}b), and the projection directions are determined by the face norm at each location. Thus at locations with very large curvatures, these non-parallel projection directions can cause overlapping.

Although not ideal, overlapped cells can be printed successfully and only happen occasionally (e.g., the model in Figure~\ref{fig:bunny}a has three pairs out of 179 cells overlap).
To eliminate cell overlapping, the tool can be improved by detecting the overlapped cells at the backend and then re-computing and shrinking cell sizes dynamically at run time. 

\subsubsection{Blank regions}

As described in section \ref{cellGen}, the tool creates a porous mesh by subtracting evenly distributed polyhedra from a mesh body through the \verb|BooleanDifference| command in Rhino3D. 
Occasionally, when the mesh surface is rugged (e.g., the connecting region between the bunny ears in Figure~\ref{fig:bunny}), the mesh fails to intersect with the polyhedron, causing a failed boolean operation and hence a blank region. 

Similar to overlapped cells, the blank spaces can be detected at the backend of the tool and further resolved computationally. For example, the blank regions can be re-processed with elongated polydedra to ensure a successful \verb|BooleanDifference| command.

\subsubsection{Cell properties}
Our current tool supports two types of cell shapes: tessellation patterns (square and hexagon), which are typical shapes used to fill a plane without gaps \cite{gullberg1997mathematics}, and circles, which are commonly found in organic porous geometries in nature \cite{tian_2020_porous}. To create more expressive displays, we can further enhance the tool by allowing users to generate their own custom cross-section shapes. For example, a designer may create or import a houndstooth shape that is commonly seen in fashion design. They may also mix different display cell shapes within the same model for expressivity. 

Additionally, the cell size and their distributions may also vary across the surface of the 3D model to present different display resolutions within the same 3D shape. For example, with the printed Stanford bunny in Figure~\ref{fig:bunny}, the hexagonal cells are uniformly distributed across the entire shell. A future version of the tool may allow the designers to select an area of interest using a lasso selection tool and fine-tune its resolution, e.g., a higher density of cells around the bunny's eyes while maintaining a coarser cell distribution on the body.

\section{Conclusion}
We presented a pipeline that allows the design and printing of custom 3D magnetophoretic displays that are self-contained, always-on, and interactive. We showed the magnetic liquid mixture selections, the 3D printer hardware modifications, and the empirical results on the properties of the  printed magnetophoretic cells. We also introduced the companion software solution, the interface and the algorithm that convert a static 3D model into a display, and the slicing G-code to automate the printing process. We concluded with a suite of examples and a discussion on the limitations as well as the future directions of 3D magnetophoretic displays.

\bibliographystyle{ACM-Reference-Format}
\bibliography{sample-base}

\end{document}